\documentclass[aps,titlepage,twocolumn,superscriptaddress]{revtex4}
\usepackage{epsfig}
\usepackage{times}

\begin{document}

\title{Seasonal payoff variations and the evolution of cooperation in social dilemmas}

\author{Attila Szolnoki}
\email{szolnoki.attila@energia.mta.hu}
\affiliation{Institute of Technical Physics and Materials Science, Centre for Energy Research, Hungarian Academy of Sciences, P.O. Box 49, H-1525 Budapest, Hungary}

\author{Matja{\v z} Perc}
\email{matjaz.perc@gmail.com}
\affiliation{Faculty of Natural Sciences and Mathematics, University of Maribor, Koro{\v s}ka cesta 160, SI-2000 Maribor, Slovenia}
\affiliation{Complexity Science Hub Vienna, Josefst\"{a}dterstra{\ss}e 39, A-1080 Vienna, Austria}

\begin{abstract}
Varying environmental conditions affect relations between interacting individuals in social dilemmas, thus affecting also the evolution of cooperation. Oftentimes these environmental variations are seasonal and can therefore be mathematically described as periodic changes. Accordingly, we here study how periodic shifts between different manifestations of social dilemmas affect cooperation. We observe a non-trivial interplay between the inherent spatiotemporal dynamics that characterizes the spreading of cooperation in a particular social dilemma type and the frequency of payoff changes. In particular, we show that periodic changes between two available games with global ordering best be fast, while periodic changes between global and local ordering games best be slow for cooperation to thrive. We also show that the frequency of periodic changes between two local ordering social dilemmas is irrelevant, because then the process is fast and simply the average cooperation level of the two is returned. The structure of the interaction network plays an important role too in that lattices promote local ordering, whilst random graphs hinder the formation of compact cooperative clusters. Conversely, for local ordering the regular structure of the interaction network is only marginally relevant as role-separating checkerboard patterns do not rely on long-range order.
\end{abstract}

\maketitle

\section{Introduction}

When only the fittest survive, cooperation is an unsustainable proposition because it entails the sacrifice of individual fitness for the benefit of others. Yet eusocial insects like ants and bees are famous for cooperating, even sacrificing their own reproduction ability to support their communities \cite{wilson_71}. Birds are also cooperating, frequently engaging in allomaternal behavior to help the offspring of others \cite{skutch_co61}. Microorganisms cooperate through the sharing of resources and forming biofilms \cite{nadell_fems09}. Not least humans have been dubbed supercooperators for our remarkable cooperative drive \cite{nowak_11}. Indeed, cooperation is the basis for the main evolutionary transitions that led from single cells to multicellular organisms, and to animal and human societies \cite{maynard_95}, and it remains a grand challenge across the social and natural sciences \cite{santos_jtb12, nowak_jtb12, perc_bs10, rand_tcs13, pacheco_plrev14, fu2017leveraging, chen2018social, horita2017reinforcement, perc_pr17}.

Research in evolutionary game theory \cite{sigmund_93, weibull_95, hofbauer_98, nowak_06, sigmund_10} has revealed now famous and thoroughly established mechanisms that can explain cooperation. They are kin selection \cite{hamilton_wd_jtb64}, direct and indirect reciprocity \cite{trivers_qrb71, axelrod_s81}, network reciprocity \cite{nowak_n92b}, and group selection \cite{wilson_ds_an77}, as reviewed in \cite{nowak_s06}. During the last decade, methods of statistical physics and network science \cite{boccaletti_pr06, holme_sr12, kivela_jcn14, javarone18} have been successfully integrated into the mainstream research concerning the evolution of cooperation, revealing that the structure of the interaction network can be crucial \cite{zimmermann_pre04, santos_prl05, gomez-gardenes_prl07, fu_pre09, du_wb_epl09, gomez-gardenes_epl11, ohdaira_jasss11, wu_zx_epl15, chen_w_pa16, liu_ph_pa17, allen_pre18, lee_hw_jcn18, yang_hx_epl18, fotouhi_rsif19, liu_dn_pa19}. It has also been thoroughly established that heterogeneity in general, for example in the form of heterogeneous networks, noisy payoff disturbances, or other individual properties like the teaching activity or the mobility to connect to additional other players, strongly promotes cooperation \cite{szolnoki_epl07, santos_n08, perc_pre08, zhu_pa14, yuan_wj_pone14, javarone_epjb16, javarone2016role, amaral2018heterogeneous2, zhang_w_pa17, richter_bs17, takesue_epl18, inaba_g19, chen_ys_pa16, cong_r_srep17}.

Heterogeneity may also manifest in varying environmental conditions, which alter the relations among interacting individuals \cite{alonso_jsm06, ashcroft_jrsif14, szolnoki_epl17}. The latter primarily affects payoffs, and it is often a realistic assumption that the environment, and hence the payoff values that describe the collective relations of players and their environment, change periodically. This is commonly referred to as seasonal effects. Evidently, if the periodic length of such effects is long enough then the population will converge to the solution that is related to actually applied payoff values over a sufficiently long period of time. However, much more interesting phenomena can be expected if the periodic length is shorter or comparable to the time scale of the inherent evolutionary dynamics. The question is whether in that case we can expect a new type of solution? For example, can we observe a cooperation level that is higher than the simple average of levels obtained for sub-solutions at different permanent payoff values? An otherwise ``inaccessible'' state may emerge when the different time scales interact, or alternatively, when frequent mutations trigger the appearance of such exotic solution \cite{kotil_nee18, tarnita_nee18}.

\begin{figure*}
\centerline{\epsfig{file=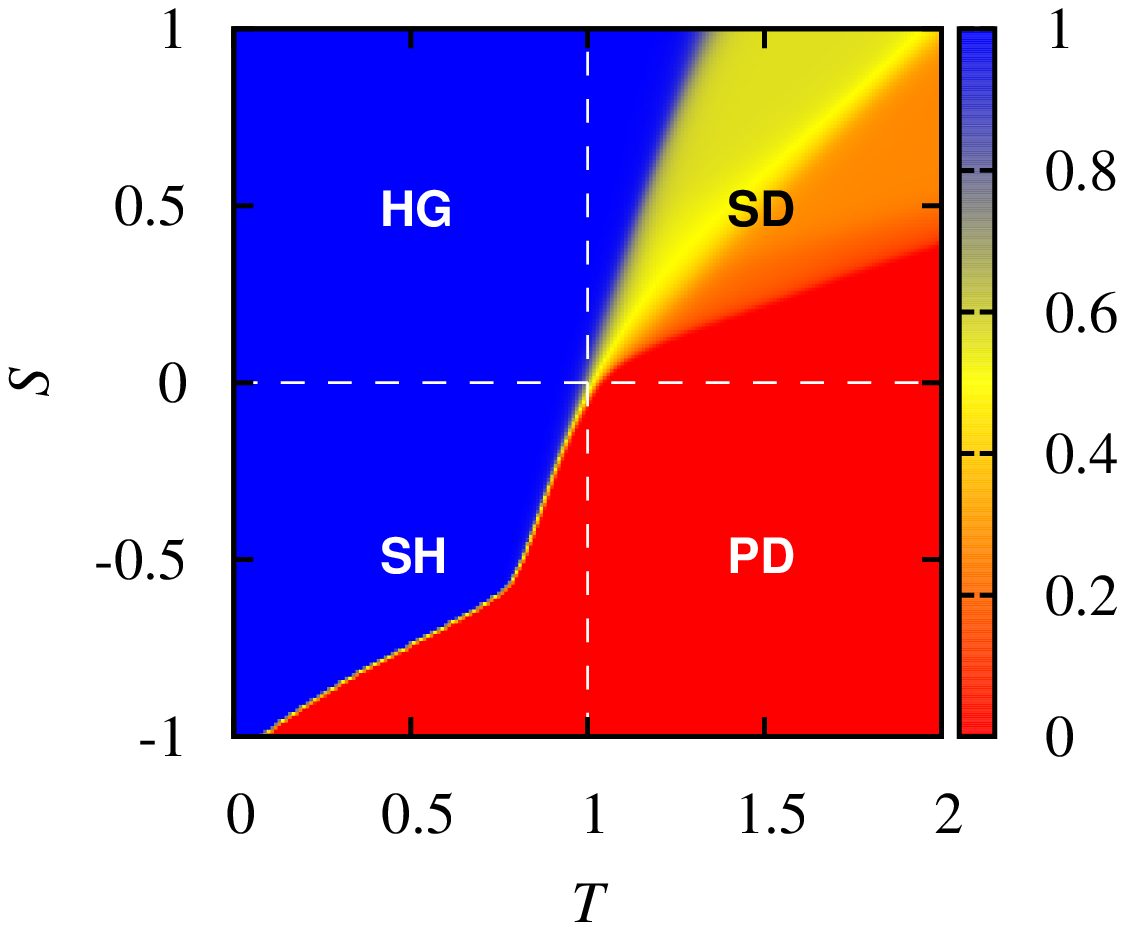,width=6.2cm}\epsfig{file=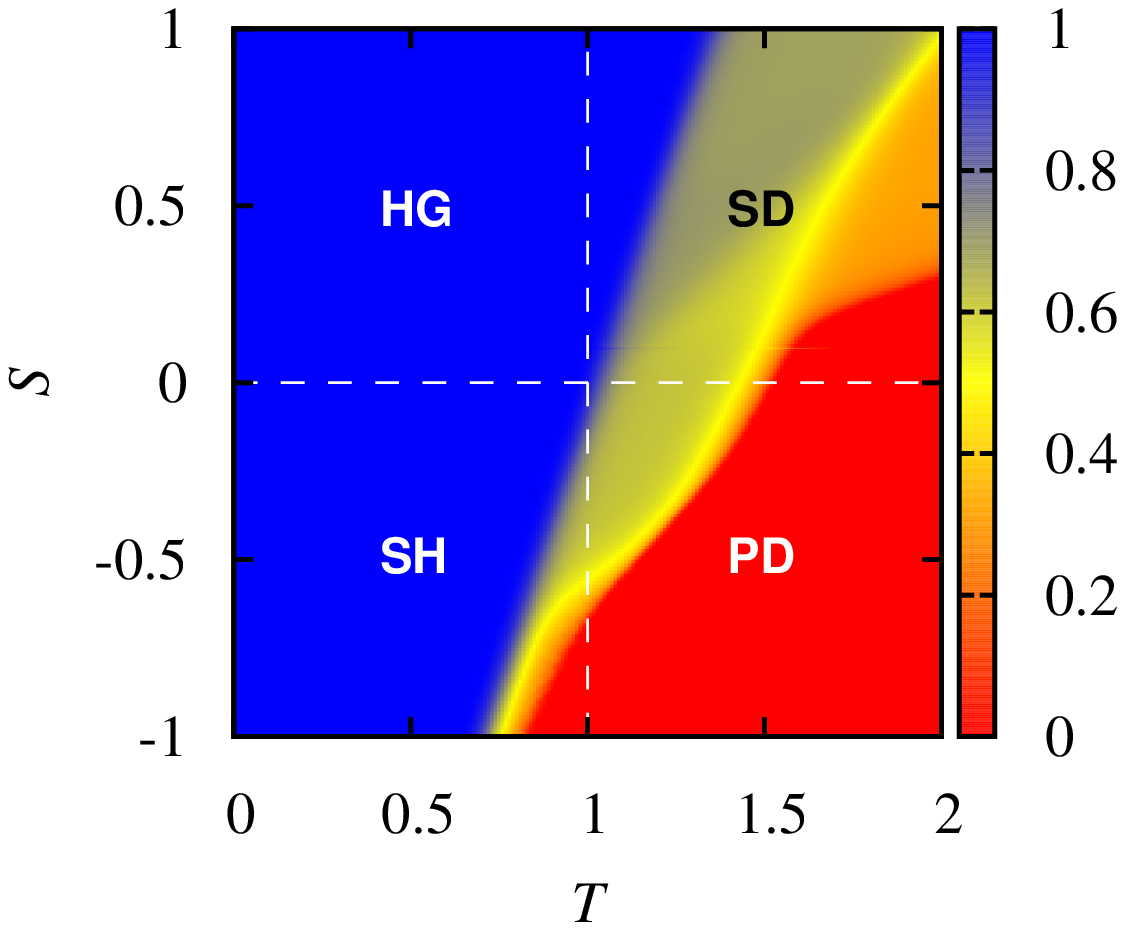,width=6.2cm}}
\centerline{\epsfig{file=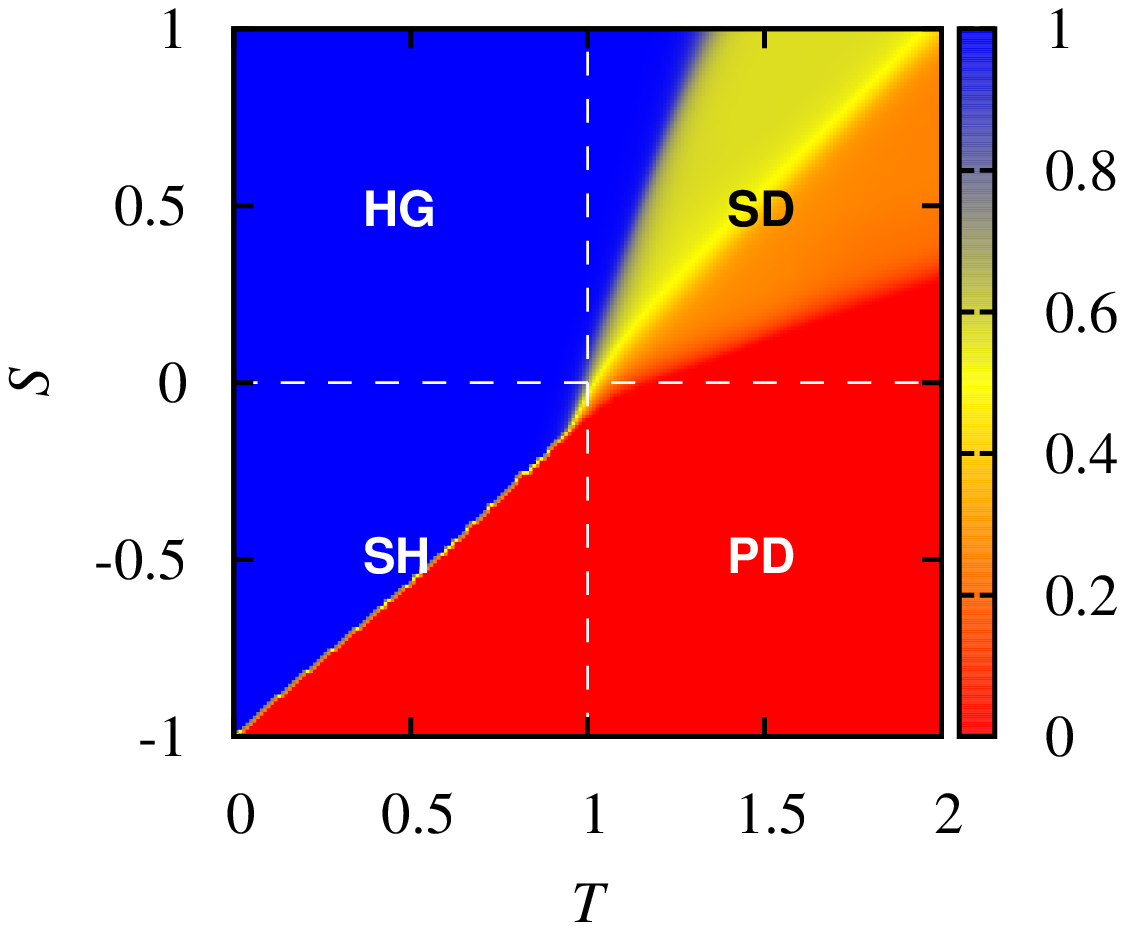,width=6.2cm}\epsfig{file=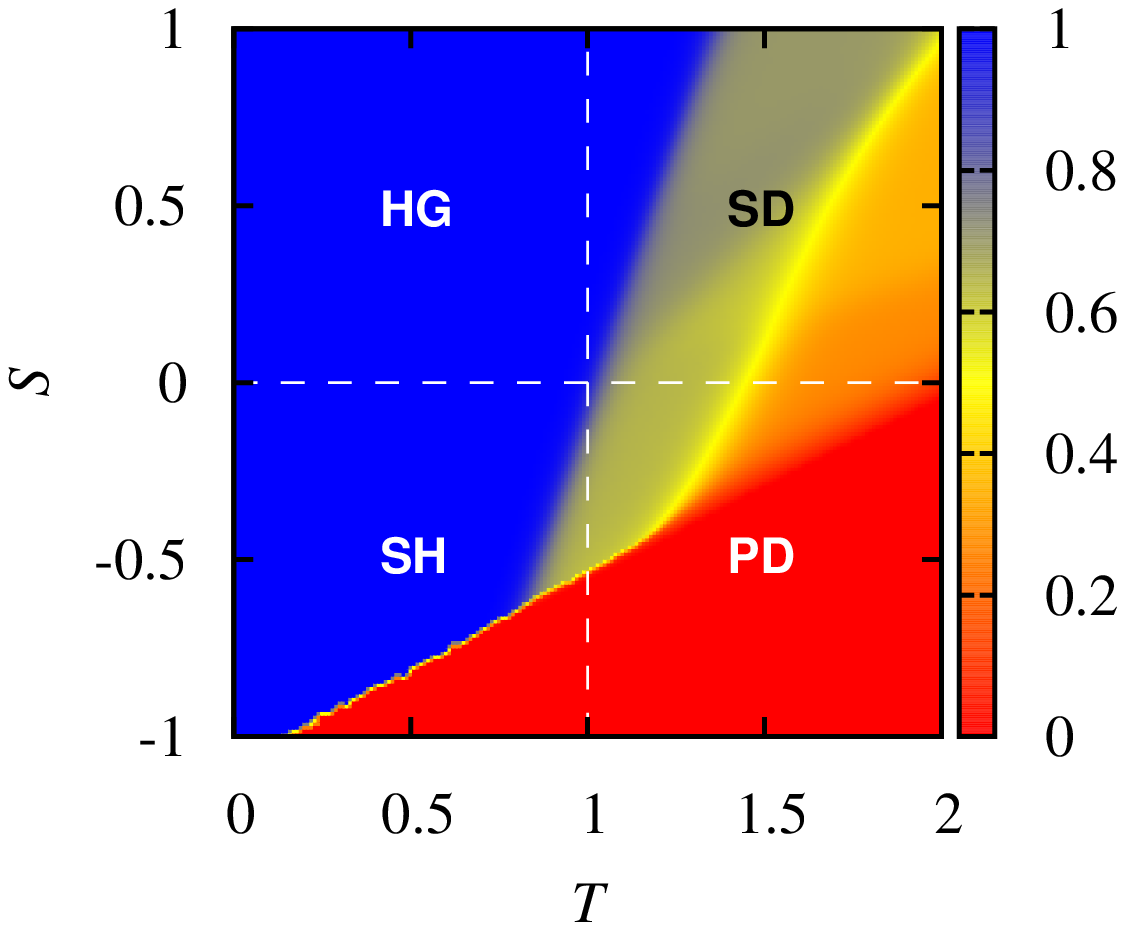,width=6.2cm}}
\caption{Color maps encoding the stationary density of cooperators on the full $T-S$ plane. Top two panels show results obtained on the square lattice, while the bottom two panels show results obtained on regular random graphs. Moreover, left two panels show results obtained for the classical version of the game where ($T_2=T, S_2=S$) are applied  permanently, while the right two panels show results obtained obtained when $(T_1=0.9,S_1=0.1)$ and $(T_2=T,S_2=S)$ are exchanged periodically with period $\tau=1$. In the snowdrift quadrant (SD), seasonal effects have a very similar impact on cooperation regardless of the applied interaction network. Conversely, in the stag-hunt (SH) and the prisoner's dilemma quadrant (PD), cooperators fare better under seasonal variations on the regular lattice than they do on regular random graphs.}
\label{TS}
\end{figure*}

In what follows, we propose a simple mathematical model that relies on the formalism of evolutionary social dilemmas (see e.g. \cite{hilbe_pnas18, danku_epl18,fu_mj_pa19}), where cooperators and defectors play a game to collect their payoffs and then either retain their strategy or imitate one of their neighbors. Seasonal payoff variations are introduced simply by means of the periodical exchange of two pairs of payoff values, which characterize the relation of competing strategies. Unlike the majority of research studying the impact of heterogeneity on the evolution of cooperation, we emphasize that in our case a pair of payoff values applies to all the players uniformly as any given time. Thus, apart from the periodic changes, no additional heterogeneity is introduced via the seasonal payoff changes. As we will show, there exists a non-trivial interplay between the inherent spatiotemporal dynamics that characterizes the spreading of cooperation in a particular social dilemma type and the frequency of payoff changes. Depending on whether the spreading of cooperation is governed by global ordering, i.e., cooperators survive of spread in compact clusters, or by local ordering, i.e., cooperators and defectors are typically arranged in role-separating mixed patterns, either fast or slow seasonal changes work best to promote cooperation, and the structure of the interaction network plays an important or relatively marginal role. Only in a special case is the impact of seasonal payoff variations relatively trivial, but in all cases a deeper understanding of the different evolutionary outcomes is attainable by studying in detailed the spatiotemporal dynamics and in particular the spreading process of cooperators.

\section{Results}

\subsection{Mathematical model}

\begin{figure}
\centerline{\epsfig{file=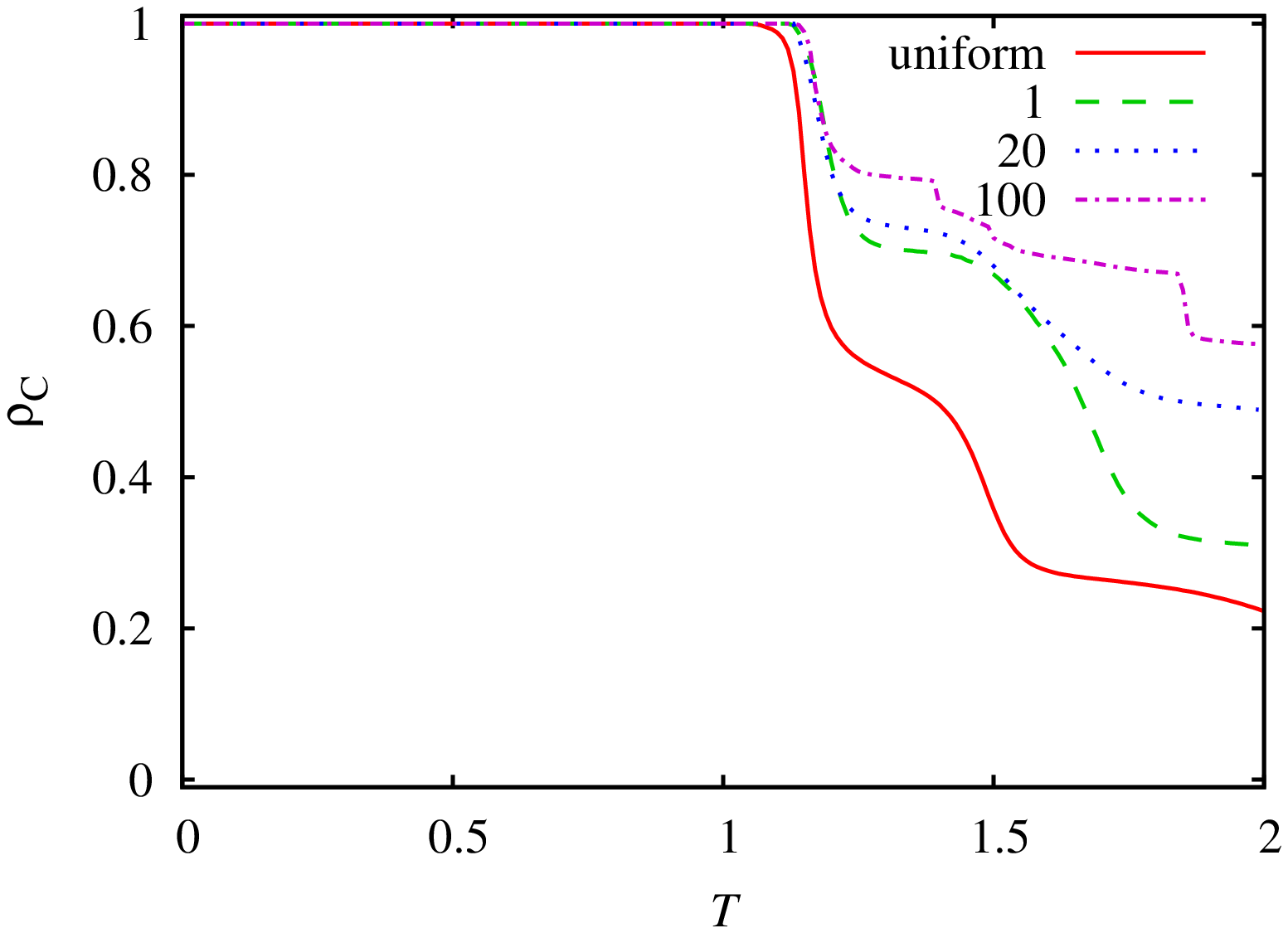,width=8.5cm}}
\centerline{\epsfig{file=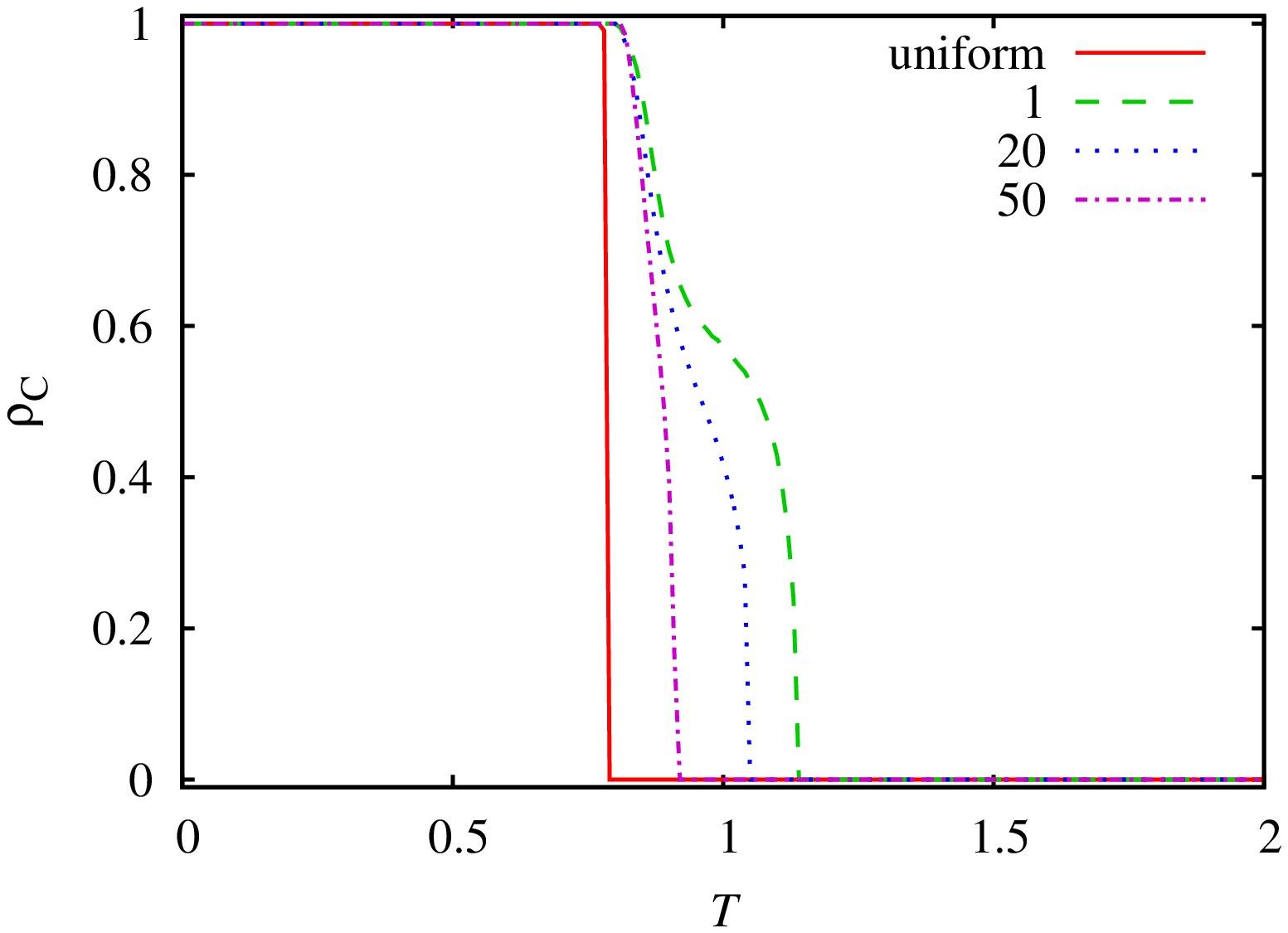,width=8.5cm}}
\caption{Stationary density of cooperators $\rho_C$ in dependence of $T_2=T$, as obtained on the square lattice using $S_2=0.5$ (left panel, cutting across the harmony and the snowdrift game) and $S_2=-0.5$ (right panel, cutting across the stag-hunt and the prisoner's dilemma game). The curves represent results obtained for different values of $\tau$, as indicated in the legend. For comparison, the evolutionary outcome obtained with the classical version of the game is also shown (uniform). It can be observed that in the snowdrift game longer periods results in a higher average cooperation level, while in the stag-hunt game more frequent seasonal changes have a similarly  positive effect (the same holds for the prisoner's dilemma game, not shown).}
\label{cross}
\end{figure}

We use the social dilemma game formalism as the basis for our model, where cooperation and defection are the two possible strategies each player can choose from. The game is played in a pairwise manner (see Methods for details). If both players decide to cooperate, they both receive the reward $R$. On the other hand, if both players decide to defect, they both receive the punishment $P$. If one cooperates while the other defects, cooperators receives the sucker's payoff $S$, while the defector receives the temptation $T$. To reduce the dimensionality of the parameter space, we set $R = 1$ and $P=0$ as fixed. In that way, the other two payoffs can have values $-1 \leq S \leq 1$ and $0 \leq T \leq 2$. If $T>R>P>S$ the social dilemma type is the prisoner's dilemma game, if $T>R>S>P$ we have the snowdrift game, $R>T>P>S$ yields the stag-hunt game, and finally if $T<P<S<R$ the payoff ranking corresponds to the harmony game (because cooperation always wins in the harmony game, formally this game is not a social dilemma). During an instance of the game a player $i$ obtains the payoff $\Pi_{i}$ in agreement with the above rules.

To introduce seasonal payoff variations, we periodically exchange two pairs of payoff values, $(T_1,S_1)$ and $(T_2,S_2)$, after $\tau$ steps. Importantly, whichever pair of payoff values applies at any given time, it does so for all the players uniformly. This setup thus differs from previously studied multigames, where different players in the population can adopt different payoff values at a particular time \cite{hashimoto_jtb14, wang_z_pre14b, szolnoki_epl14b}.

Unless stated otherwise, we predominantly use $T_1=0.9$ and $S_1=0.1$, while $T_2=T$ and $S_2=S$ are free parameters. In this way we have a cooperation supporting harmony game period, while in the other half of the time players play a proper social dilemma game that is defined by the actual values of $T$ and $S$, which are thus the only two free parameters.

\subsection{Evolution of cooperation}

\begin{figure}
\centerline{\epsfig{file=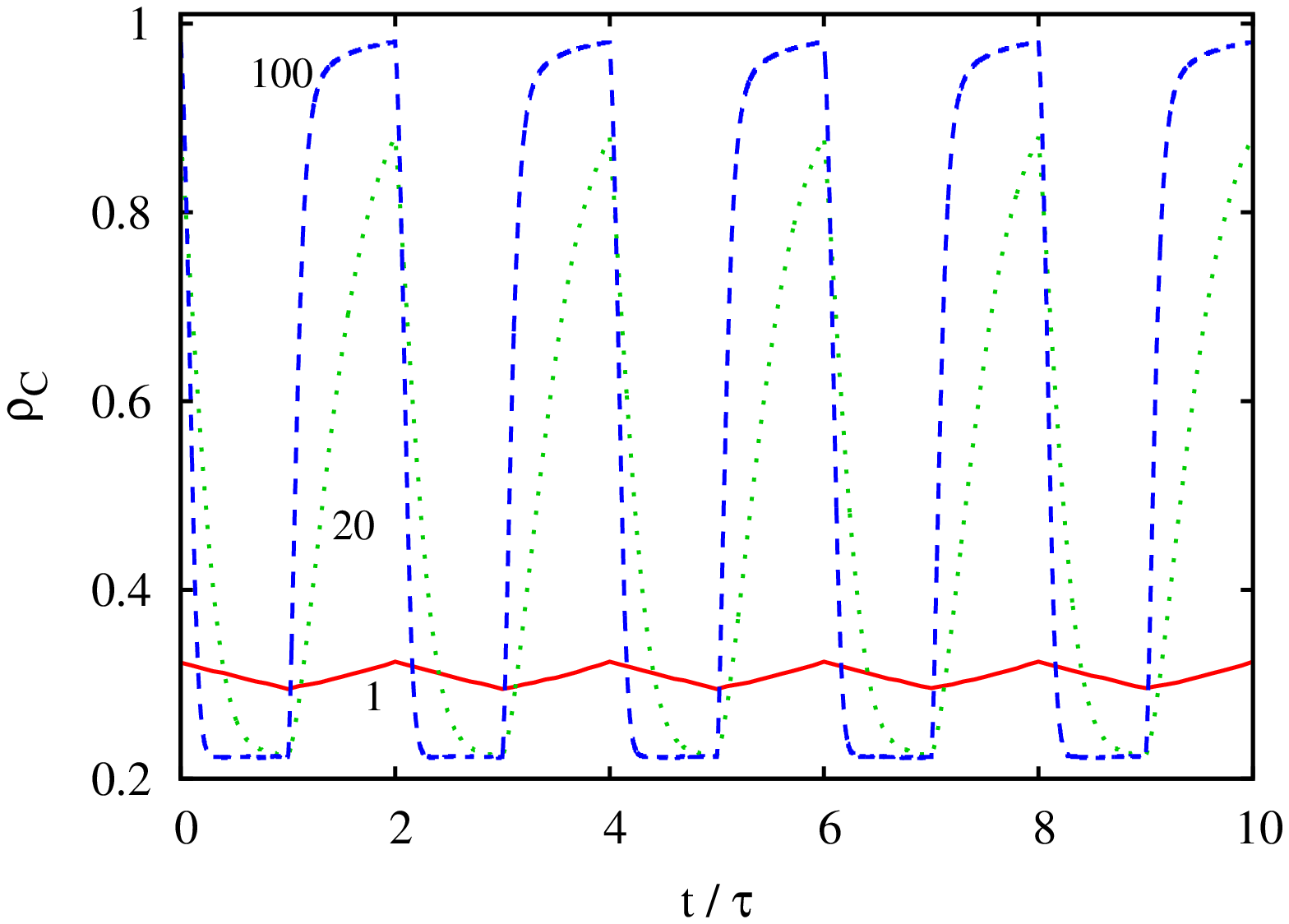,width=8cm}}
\centerline{\epsfig{file=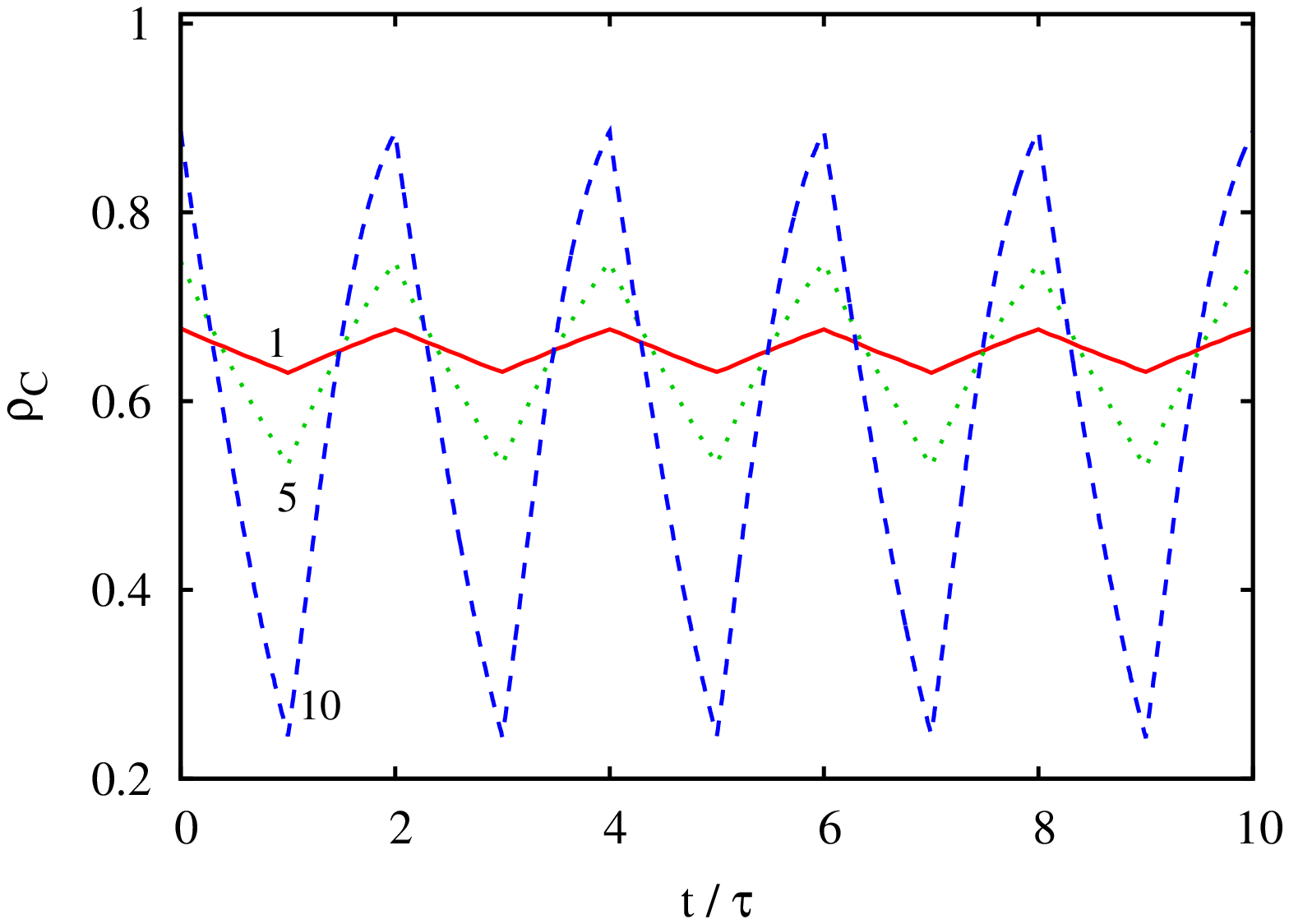,width=8cm}}
\caption{Time evolution of the cooperation level $f_C$ in the stationary state, as obtained for different values of $\tau$ in the snowdrift game ($T=2$, $S=0.5$, left panel) and in the stag-hunt game ($T=0.92$, $S=-0.5$, right panel). The applied values of $\tau$ are indicated alongside the corresponding lines. For a proper comparison we have used a normalized time scale in $1/\tau$ units. It can be observed that the relaxation dynamics for the snowdrift game is fast, while for the harmony game and the stag-hunt game, it is comparatively much slower. This is due to the fundamentally different spatiotemporal dynamics, which in the former case is governed by locally ordered role-separating checkerboard patterns, while in the later case it is governed by globally ordered compact clusters.}
\label{time}
\end{figure}

In Fig.~\ref{TS}, we show color maps that encode the long-time average fraction of cooperators, obtained when we apply periodically changing payoff values using $\tau=1$ as the periodic time. The top row shows results obtained on the square lattice, while the bottom row shows results obtained on regular random graphs where each player has the same degree $k=4$ as on the lattice [the same degree is retained to allow a direct and relevant comparison (cf. \cite{szolnoki_pre11c})]. Moreover, the left two panels show the stationary fraction of cooperators as obtained when $T_2=T$ and $S_2=S$ are applied  permanently, whereas the right two panels show the stationary fraction of cooperators as obtained when $(T_1=0.9,S_1=0.1)$ and $(T_2=T,S_2=S)$ are exchanged periodically with period $\tau=1$. Two generally valid observations are as follows. Firstly, seasonal effects have a very similar impact on cooperation in the snowdrift quadrant (SD) independently of the applied interaction network. Secondly, in the stag-hunt (SH) and the prisoner's dilemma (PD) quadrant, the cooperators fare better on the regular lattice as they do on regular random graphs. In particular, in the stag-hunt quadrant the full cooperation state occupies a larger area of the parameter plane on the regular lattice.

\begin{figure*}
\centerline{\epsfig{file=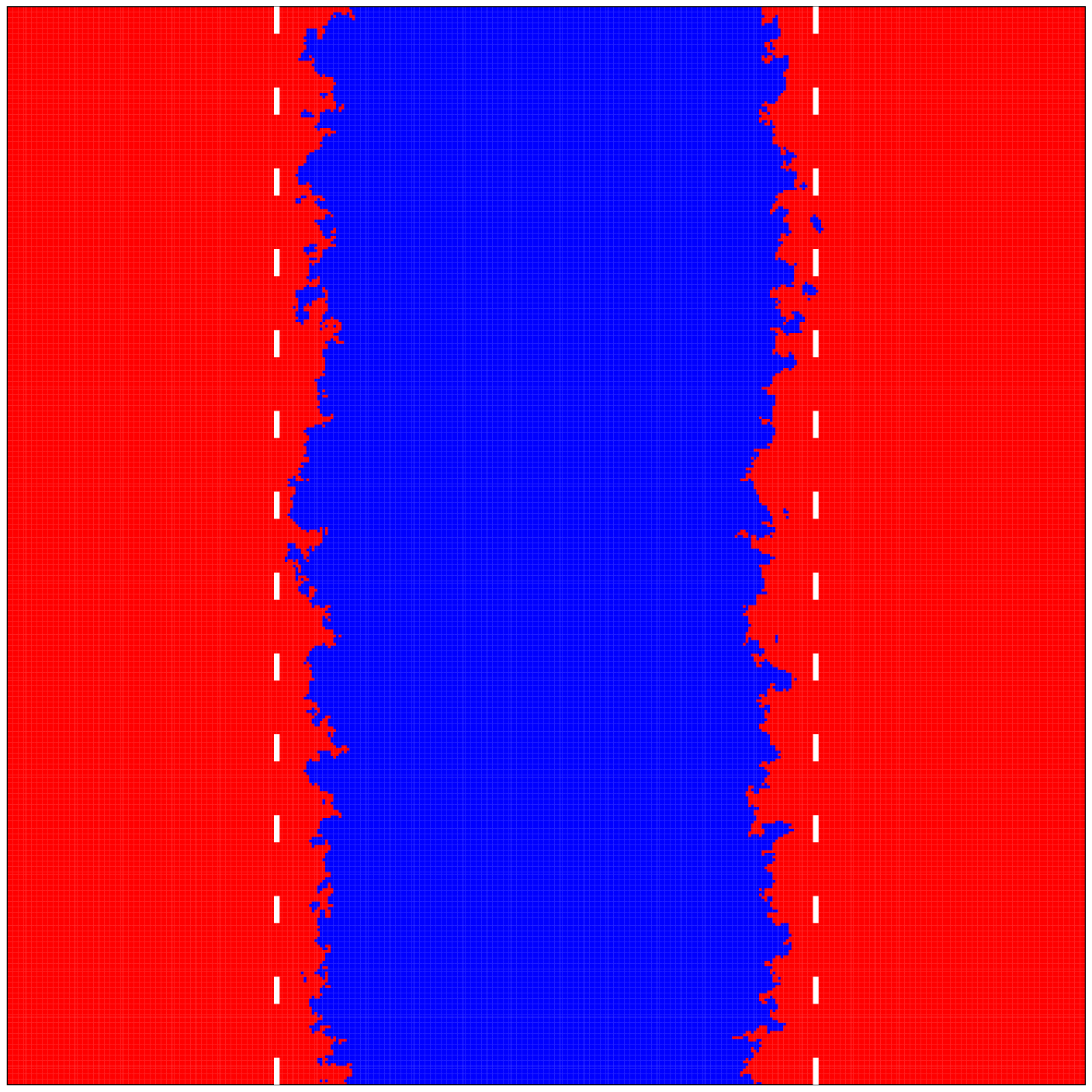,width=5.33cm}\epsfig{file=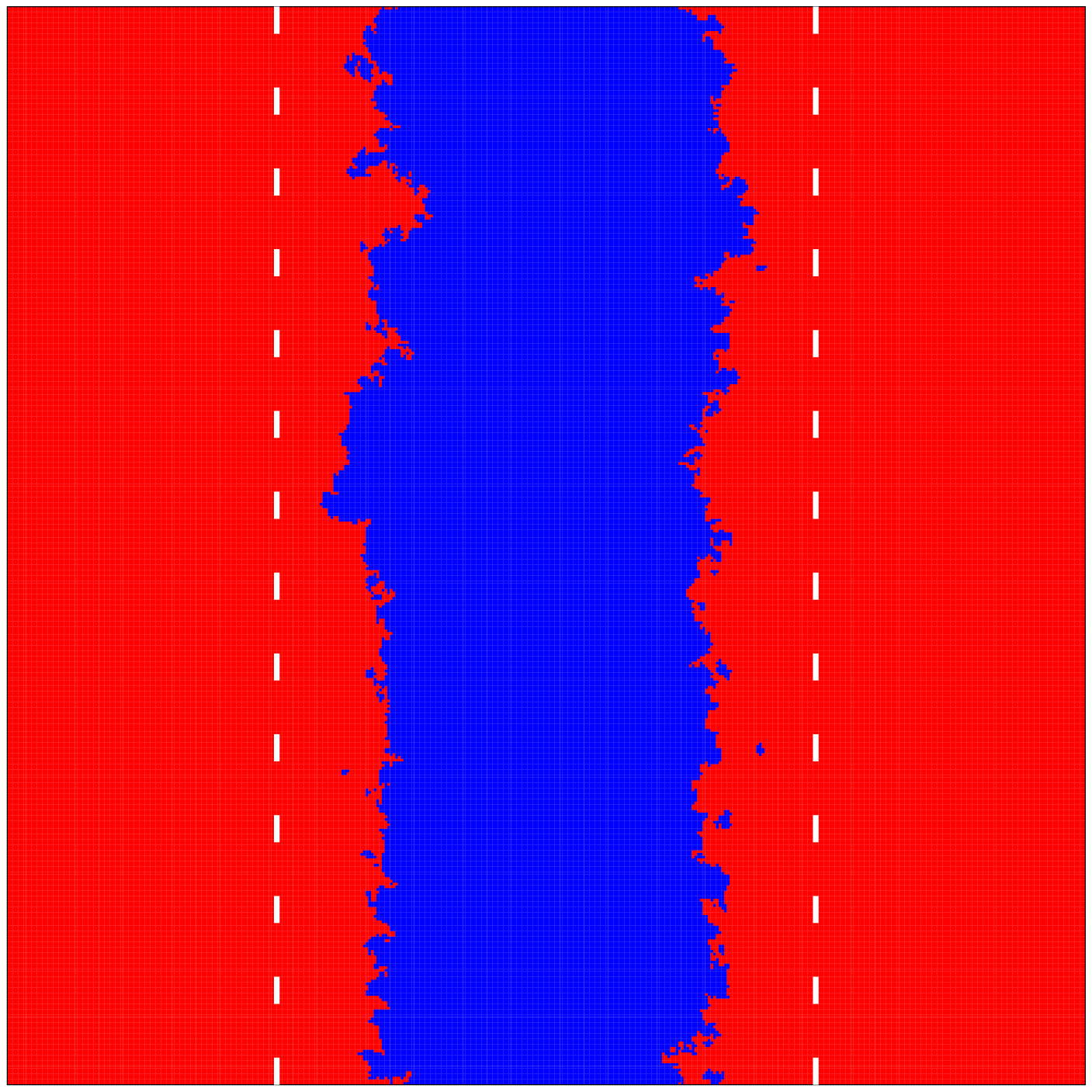,width=5.33cm}\epsfig{file=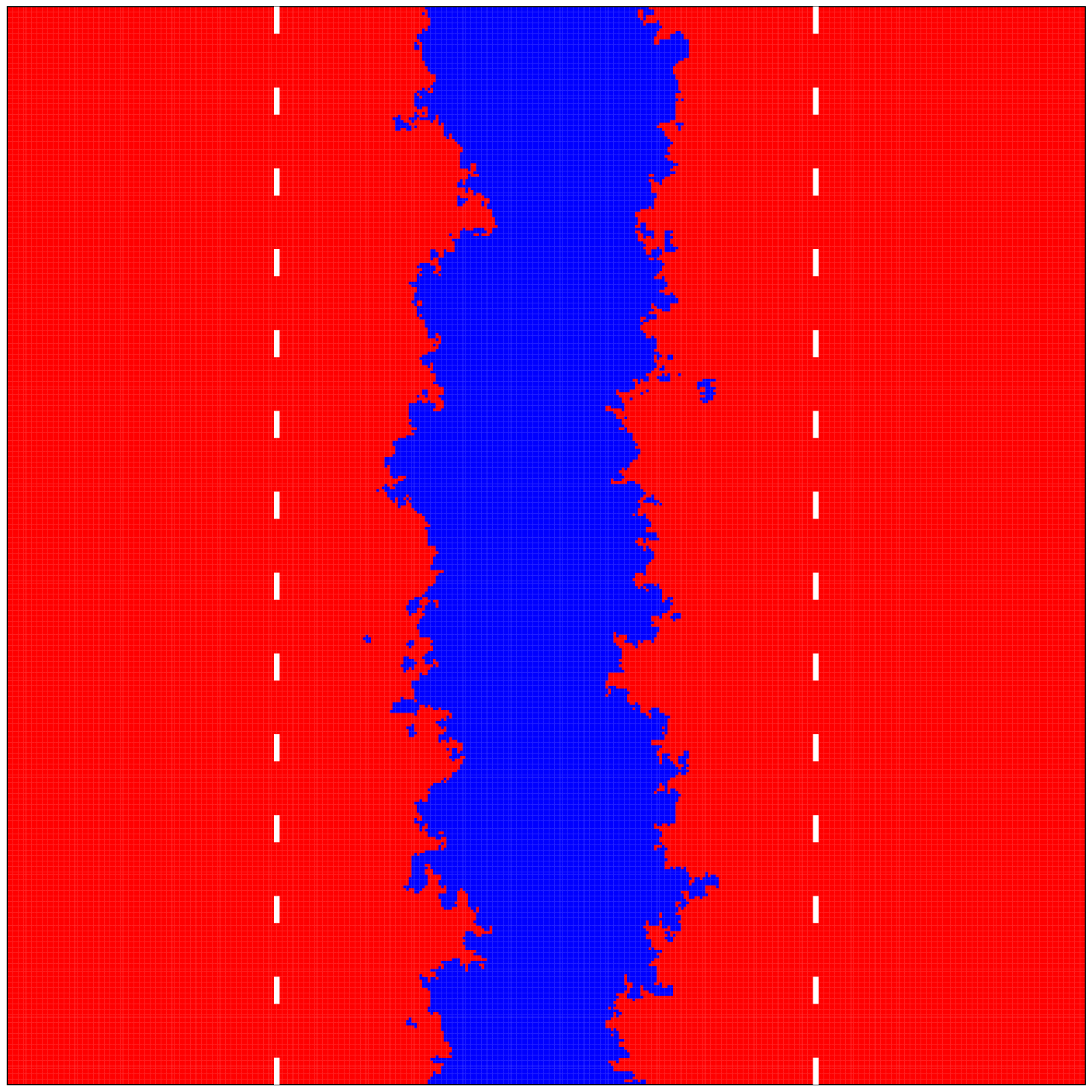,width=5.33cm}}
\caption{The propagation of defector fronts in the stag-hunt game, as obtained for $T_2=0.92$, $S_2=-0.5$, and $\tau=100$. The evolution starts from a horizontal strip od blue cooperators, marked by white dashed lines. After 200 (left), 400 (middle), and 600 (right) steps the original strip shrinks gradually because the invasion of defectors during the ($T_2,S_2$) period is more effective than the invasion of cooperators during the ($T_1=0.9,S_1=0.1$) period. We here used $L=400$ linear size for clarity, as in the corresponding animations obtained with smaller values of $\tau$ \cite{sh_A1,sh_A20,sh_A50}.}
\label{shrink}
\end{figure*}

\begin{figure}
\centerline{\epsfig{file=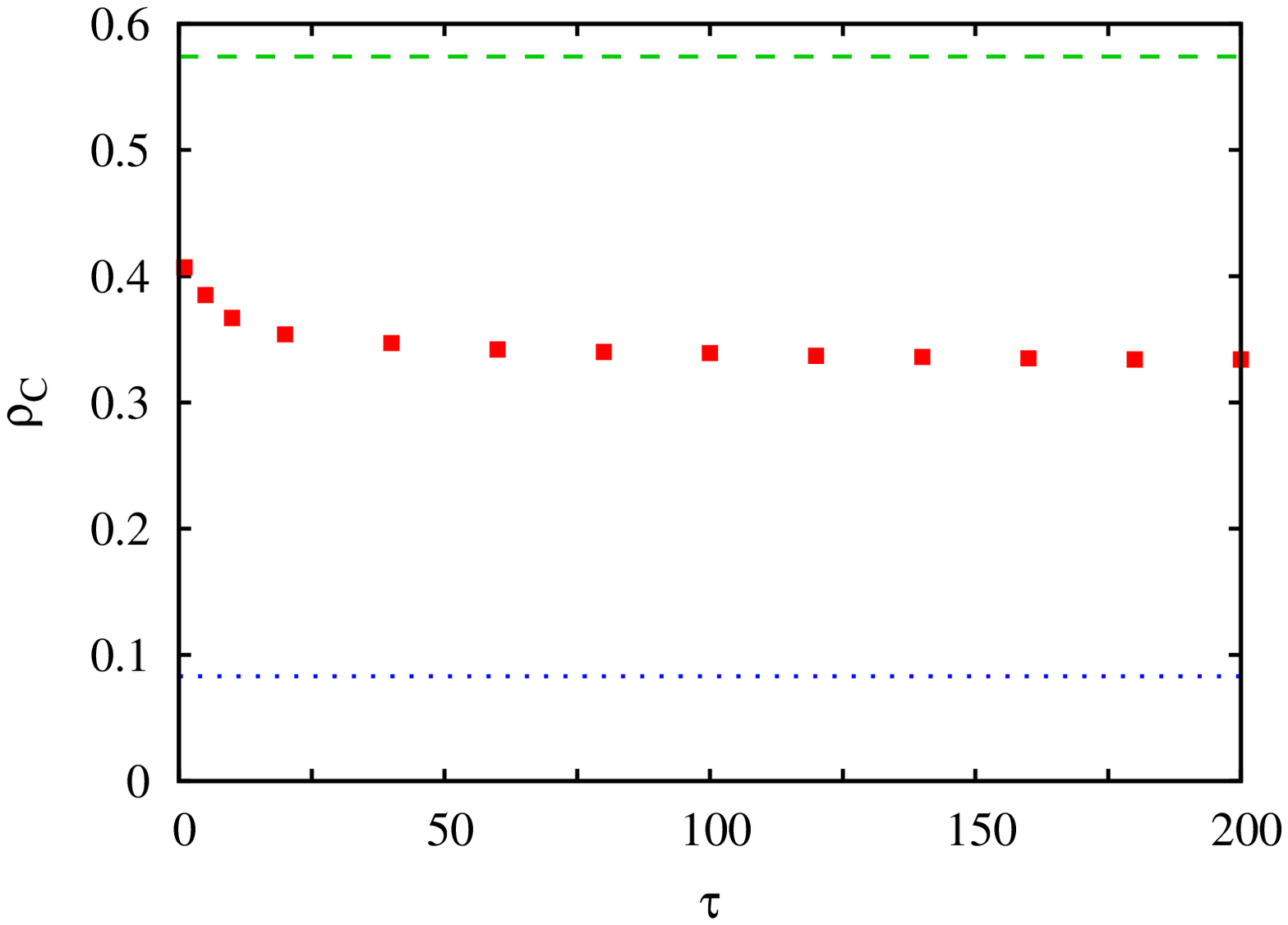,width=8cm}}
\centerline{\epsfig{file=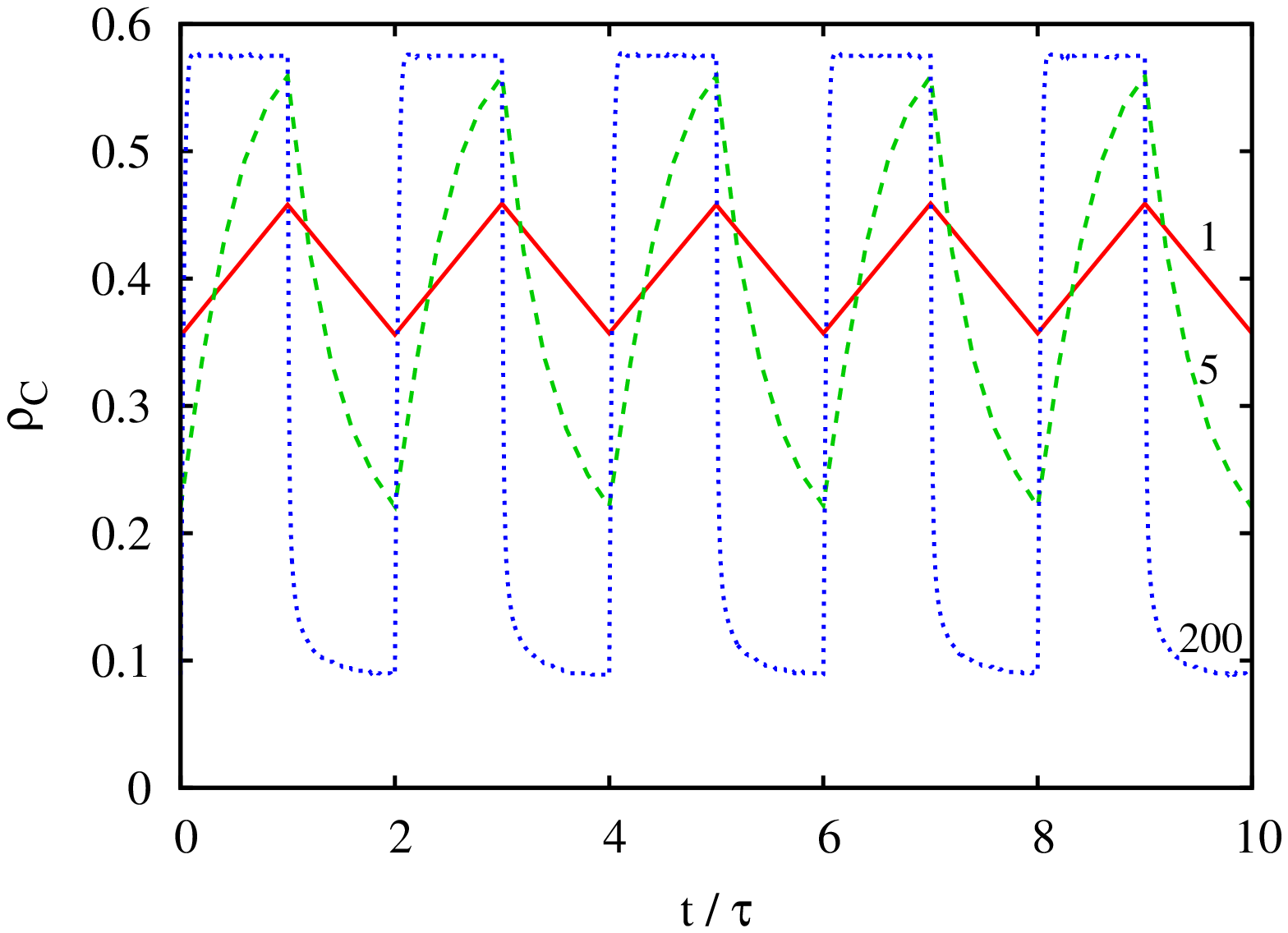,width=8cm}}
\caption{Stationary density of cooperators $\rho_C$ in dependence on the periodic time $\tau$, obtained when both payoff pairs fall into the snowdrift quadrant (left panel). We have used ($T_1=1.3$, $S_1=0.7$) and ($T_2=1.7, S_2=0.3$), whereby the corresponding $\rho_C$ values (if either payoff pair would be applied permanently) are marked by horizontal dashed lines for comparison. It can be observed that the $\rho_C$ value of the game with seasonal variations is more or less exactly the average of the two games that are alternated, and this regardless of the value of $\tau$. Only when $\tau$ is very small is the relaxation in both cases not fast enough and so the average diverges towards the snowdrift game with a slightly faster relaxation [governed by ($T_1$, $S_1$), more supportive of cooperation, marked by the green dashed line]. The right panel shows the time evolution of the cooperation level $f_C$ in the stationary state, as obtained for different values of $\tau$ using the same payoff pairs as in the left panel. The applied values of $\tau$ are indicated alongside the corresponding lines. For a proper comparison we have used a normalized time scale in $1/\tau$ units. It can be observed that the relaxation dynamics is always (in both periods) relatively fast, indicative of the fast emerging locally ordered role-separating checkerboard patterns that govern both instances of the applied snowdrift game.}
\label{sd_sd}
\end{figure}

To understand the latter diverse impact of different interaction networks for different social dilemmas, we first show in Fig.~\ref{cross} a cross section of the $T-S$ plane for the square lattice using $S=0.5$ (left panel, cutting across the harmony and the snowdrift game) and $S=-0.5$ (right panel, cutting across the stag-hunt and the prisoner's dilemma game), as obtained for different values of $\tau$. These results reveal that the application of longer periods helps increasing the cooperation level in the snowdrift game, while conversely, for the stag-hunt game the shorter the periods the higher the level of cooperation. The same conclusion holds for the prisoner's dilemma game, which would be visible if a somewhat less negative value of $S$ is used (not shown). As we will show in what follows, these differences are routed in the fundamentally different spreading processes that govern the spatiotemporal evolutionary dynamics. In particular, while for the snowdrift game fast emerging role-separating checkerboard patterns dominate \cite{szabo_jtb12, amaral2017role}, for the stag-hunt and the prisoner's dilemma game cooperation proliferates by means of compact clusters, thus requiring longer relaxation times due to the need of a globally ordered state.

In support of this explanation, we show in Fig.~\ref{time} the time dependence of the cooperation level $f_C$ for a specific snowdrift game parameterization, as obtained for different values of $\tau$. The left panel clearly shows that $f_C$ reaches the minimal value that corresponds to the ($T_2,S_2$) payoff pair very fast, even for relatively small $\tau$ values. The maximal $f_C$ value, which would correspond to the ($T_1,S_1$) payoff pair, however, is never reached even if the value of $\tau$ is very large. This confirms that the role-separating spatially mixed strategy patterns that characterize the snowdrift game emerge fast, while the emergence of compact clusters driven by a globally ordered state in the stag-hunt game takes much longer. Consequently, the positive impact of the harmony game can only manifest itself in the average level of cooperation $\rho_C$ as the period $\tau$ becomes sufficiently long -- indeed, the longer the better, as shown in the left panel of Fig.~\ref{cross}.

The described asymmetric ordering speed can be demonstrated clearly with animations, where we show the time evolution for $\tau=1$ \cite{sd_A1}, $\tau=20$ \cite{sd_A20}, and for $\tau=100$ \cite{sd_A100} by using the same $(T_2=2,S_2=0.5)$ pairs. The movies show that the ``mixed'', or role-separating order emerges almost immediately for all $\tau$ values, while the homogeneous cooperative (blue) domain can hardly evolve because isolated defectors prevent reaching the absorbing full $C$ state. The same argument also explains why we observe very similar behavior for regular random graphs in the snowdrift quadrant. In fact, the role-separating ordering is not hindered by long-range links, and so the only decisive factor in this case if the type of the social dilemma, while the topology of the interaction network has only marginal importance. At a closer inspection, the only inferrable difference in the snowdrift quadrant in Fig.~\ref{TS} between the square lattice and regular random graphs can be observed for $1.5<T<2$ and $0<S<0.3$, where the level of cooperation is a bit higher in the latter case. This is because the global ordering that is relevant for the harmony game can emerge a bit faster due to the shortcuts, and thus somewhat raise the average level of cooperation in comparison to the square lattice, where the same process unfolds somewhat slower.

We meet with a qualitatively different outcome in the stag-hunt quadrant, because here two homogeneous states, namely full cooperation and full defection represent the permanent solutions of the two periodically alternating payoff pairs. Consequently, both extreme states require global ordering, hence requiring a kind of coordination of akin players. This explains why the time evolution of $\rho_C$ curves is more symmetric in the right panel of Fig.~\ref{time}, where we present results for the stag-hunt game. There is, however, a slight difference between how cooperator domains and defector domains grow when the actual payoff values allow their spreading. In particular, the spreading of defectors is more aggressive while the spreading of cooperators requires more coordination and smooth interfaces. The resulting difference of invasion speeds becomes visible at larger values of $\tau$, which ultimately results in a lower average cooperation level as $\tau$ increases, as shown in the right panel of Fig.~\ref{cross}.

For this case we also provide corresponding animations using $\tau=1$ \cite{sh_A1}, $\tau=20$ \cite{sh_A20} and $\tau=50$ \cite{sh_A50}, where $(T_2=0.92,S_2=-0.5)$ values are used for all cases. To illustrate the asymmetric invasion speeds more clearly, we here use alternative prepared initial conditions, where ordered states are separated by vertical borders. Especially for $\tau=50$ it can be observed nicely that the bulk of the defector phase remains practically intact while the cooperator phase first falls apart and then completely dies out. We also show this phenomenon with snapshots in Fig.~\ref{shrink} for $\tau=100$. The difference of invasion speeds of defectors and cooperators in the two governing social dilemmas also explains why the usage or regular random graphs only modestly supports cooperation in the stag-hunt quadrant, even if the periodic changes are fast and the advantages of more effective spreading of defectors ought to remain hidden. Namely, the random topology hinders the formation of compact cooperator domains with a smooth interface, and thus is not supportive of network reciprocity. At the same time, shortcuts favor the effective invasions of defectors into the cooperative domains, which taken together significantly lower the positive consequences of periodically changing payoff values in this case.

To conclude, we lastly consider the third qualitatively different case when two solutions that are both characterized by locally ordered states compete in an alternating manner. This can be reached if both $(T_1,S_1)$ and $(T_2,S_2)$ payoff pairs are from the snowdrift quadrant. Accordingly, we have chosen $(T_1=1.3, S_1=0.7)$ and $(T_2=1.7, S_2=0.3)$, where the permanent usage of one of these pairs would yield $\rho_C=0.574$ and $\rho_C=0.083$ in the stationary state, respectively. The average cooperation level in dependence on $\tau$ is shown in the left panel of Fig.~\ref{sd_sd}. As expected based on the governing fast spatiotemporal evolutionary dynamics in both cases, it can be observed that almost irrespective of the value of $\tau$ the cooperation level is simply the average of the cooperation level in the two alternating snowdrift games. The validity of this argument is confirmed by the symmetric time evolution shown in the right panel of Fig.~\ref{sd_sd} for virtually all $\tau$ values, apart for $\tau<25$ where a slight breaking of symmetry occurs. This marginal effect, however, is related to the fact that the $(T_1,S_1)$ payoff pair (depicted with the green dashed line) is closer to the randomly mixed state than the $(T_2,S_2)$ payoff pair (depicted with the blue dotted line), and therefore the relaxation is slightly faster in the former case. To sum up, only in this particular case do periodic variations in payoffs amount to the most naive expectation that the final outcome is a simple average of the two extreme cases obtained under would-be permanent conditions.

\section{Discussion}

Motivated by the fact that varying environmental conditions affect cooperation in social dilemmas, we have studied how seasonal variations, in particular periodic switching between two different types of games, affects the evolution of cooperation in structured populations. Naively, one could expect a resulting cooperation level that is simply an average of the stationary cooperation levels of the two social dilemmas entailed in the switch, but this turns out to be the case only in one special case. Namely, we have shown that if locally ordered states govern the evolution of cooperation, as is the case in the snowdrift game, then relaxation is fast enough and practically independent of the structure of the interaction network, in which case indeed the cooperation level is just a simple average of the cooperation levels in the two considered games. And this is true regardless of the structure of the interaction network, and regardless of the frequency of payoff variations.

In general, however, we have shown that there exists a non-trivial interplay between the inherent spatiotemporal dynamics that characterizes the spreading of cooperation in a particular social dilemma type and the frequency of payoff changes. The inherent spatiotemporal dynamics is affected by the parametrization of the two games entailed in the switch, and by the structure of the interaction network. More precisely, when cooperation proliferates by means of compact clusters, as is the case in the harmony game, the stag-hunt game, and the prisoner's dilemma game, the relaxation times are longer and the translation invariant feature of an interaction graph promotes the local emergence of a coordinated state that is necessary for a spreading process to reach a global ordering. Accordingly, we have shown that periodic changes between two games with global ordering should be fast for cooperation to be promoted, and preferably unfold on a network without long-range links, like a square lattice. If an interaction network with long-range links is used, like a regular random graph, then the irregular structure hinders the formation of compact clusters, which in turn impairs the evolution of cooperation. Since the observed behavior is strongly related to the diverse speed of spreading at different payoff values, we may expect conceptually similar behavior in scale-free and multilayer complex networks as well \cite{wang_z_epjb15, gleeson_prx16, arruda_pr18}.

If the seasonal changes trigger global and local ordering towards particular Nash-equilibrium, for example switching between the harmony game and the snowdrift game, then we have shown that it is best for these changes to be slow for cooperation to be promoted best. This goes on the account of the harmony game, where the growth of compact cooperative domains takes time, but is then quickly destroyed by the role-separating checkerboard patterns that govern the evolutionary dynamics in the snowdrift game.

Taken together, we have addressed fundamental questions behind the evolution of cooperation in social dilemmas under seasonal payoff variations. In particular, we have revealed the delicate interplay between the inherent spatiotemporal game dynamics and the frequency of seasonal change, and we have shown when, and under which conditions, new types of solutions can be expected that allow cooperators to survive under adverse conditions or at higher stationary densities. We hope that our exploration will be useful for further improving the theoretical understanding of the evolution of cooperation in social dilemmas, and that it will motivate further research along these lines.

\section{Methods}

\begin{figure}
\centerline{\epsfig{file=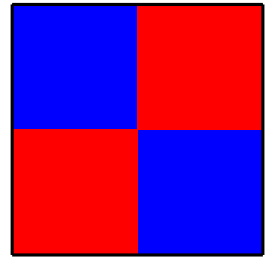,width=3.5cm}}
\caption{We use prepared initial conditions as shown, so that the population cannot evolve into a homogeneous absorbing state during the first period of $\tau$ even under the most adverse conditions. For this to hold, the linear size of each square should be more than four times larger than $\tau$. Blue color depicts cooperators $(s_i=C)$, while the red color depicts defectors $(s_i=D)$. Both strategies have equal occupance of the population at the start of evolution.}
\label{init}
\end{figure}

We have studied outcomes of the proposed evolutionary model on a square lattice of size $N=L^2$ with the von Neumann neighborhood and periodic boundary conditions, and for comparison on regular random graphs where each player has the same degree $k=4$. The square lattice is the simplest interaction network where the interaction range of each player is limited to the four nearest neighbors -- a tradition that is due to Nowak and May \cite{nowak_n92b}, which has since emerged as an often used setup to reveal all relevantly different evolutionary outcomes that are attainable in structured populations \cite{perc_pr17}. Similarly, the regular random graph can be considered as the simplest interaction network with long-range connections, whilst preserving the degree of each player so that the results remain comparable to the results obtained on the square lattice. Moreover, by retaining the degree of each player, we can study specifically the impact of long-range links (or shortcuts), without introducing further heterogeneity that would otherwise of course also affect the evolution of cooperation \cite{szolnoki_epl07, santos_n08, perc_pre08, zhu_pa14, li_j_csf18, yang_hx_pa19}.

Due to the application of periodic switching between two different parameterizations of a social dilemma, it is technically not irrelevant which parameterization is applied first because it may be that one strategy dies out before the other parameterization comes into play. This would be particularly likely if we launch the evolution from a random state and use a large $\tau$ value (long period between switching) under adverse conditions for cooperation. To avoid this, we use prepared initial conditions as shown in Fig.~\ref{init}, such that the population cannot evolve into a homogeneous absorbing state during the first period. This in turn also means that to avoid finite-size effects we have to apply sufficiently large system sizes that depend on the actual value of $\tau$. Practically, if the linear size of an initial domain exceeds twice the value of $\tau$ then this domain cannot die out during $\tau$ steps even if it shrinks at every iteration step. We note, however, that the prepared initial state still ensure equal representation of cooperators and defectors in the population at the start of the evolution.

After the application of prepared initial conditions, we use the Monte Carlo simulation method with the following three elementary steps. This is a standard procedure that has been described in detail many time before, for example in \cite{perc_bs10}. In what follows, we repeat the description for completeness of this paper. Firstly, a randomly selected player $i$ acquires its payoff $\Pi_{i}$ by playing the game with all its four neighbors. Secondly, one randomly chosen neighbor of player $i$, denoted by $j$, also acquires its payoff $\Pi_{j}$ by playing the game with all its four neighbors. Finally, player $i$ adopts the strategy $s_j$ from player $j$ with the probability
\begin{equation}
W = \frac{1}{1+\exp[(\Pi_i-\Pi_j)/K]} \, ,
\end{equation}
where $K$ quantifies the uncertainty by strategy adoptions \cite{szabo_pre98}. In the $K \to 0$ limit, player $i$ copies the strategy of player $j$ if and only if $\Pi_j > \Pi_i$. Conversely, in the $K \to \infty$ limit, payoffs seize to matter and strategies change as per flip of a coin. Between these two extremes players with a higher payoff will be readily imitated, although under-performing strategies may also be adopted, for example due to errors in the decision making or imperfect information. Without loss of generality we have here used $K=0.1$. We stress, however, that qualitatively similar behavior can be observed for other finite values of $K$. Repeating the above three elementary steps $L^2$ times constitutes one full Monte Carlo step, which thus gives a chance to every player to change its strategy once on average.

Presented results were obtained on square lattices with linear size ranging from $L=400$ to $L=1000$, and on regular random graphs with size ranging from $N=10^5$ to $N=10^6$, to avoid finite size effects. All simulations were run to obtain dynamical behavior that is independent on the applied system size, hence we can exclude finite-size effects.

\begin{acknowledgments}
This research was supported by the Hungarian National Research Fund (Grant K-120785) and the Slovenian Research Agency (Grants J4-9302, J1-9112 and P1-0403).
\end{acknowledgments}

\end{document}